# Measuring the Monetary Value of Online Volunteer Work

**Hanlin Li[1], Brent Hecht[1], Stevie Chancellor[2]**

[1] Northwestern University
[2] University of Minnesota Twin Cities
lihanlin@u.northwestern.edu, bhecht@northwestern.edu, steviec@umn.edu

## Abstract

Online volunteers are a crucial labor force that keeps many for-profit systems afloat (e.g. social media platforms and online review sites). Despite their substantial role in upholding highly valuable technological systems, online volunteers have no way of knowing the value of their work. This paper uses content moderation as a case study and measures its monetary value to make apparent volunteer labor's value. Using a novel dataset of private logs generated by moderators, we use linear mixed-effect regression and estimate that Reddit moderators worked a minimum of 466 hours per day in 2020. These hours amount to 3.4 million USD a year based on the median hourly wage for comparable content moderation services in the U.S. We discuss how this information may inform pathways to alleviate the one-sided relationship between technology companies and online volunteers.

## Introduction

Online volunteer work underpins some of the greatest technological innovations and advances in recent computing history. In addition to non-profit and open-source initiatives such as Wikipedia and Linux, online volunteer work also supports highly valued technologies such as Stack Overflow (a question-and-answer website sold for 1.8 billion in 2021). Similarly, social media platforms such as Facebook Groups, Reddit, and Discord prominently depend on fleets of volunteer moderators to build and manage communities with millions of users and, thereby, keep these platforms viable (Gilbert, 2020; Matias, 2019).

While many volunteer-supported, for-profit technologies achieve great financial success, they set up inequitable power structures in the technology sector. Online volunteers who provide the crucial labor supporting these companies are subject to worsening working conditions (Matias, 2016), monetization without consent (Arrieta-Ibarra et al., 2018; Li et al., 2019; Vincent et al., 2021), and potentially exploitation (Terranova, 2000). More broadly, online volunteers often have little power to shape the technology they co-create with for-profit companies (Vincent et al., 2021). From a policymaking perspective, online volunteer work creates new labor mechanisms by subsidizing actual compensated labor (Postigo, 2009), and scholars have suggested that companies profiting from this free work may be contributing to an industry-wide decline in labor share (the proportion of business income allocated to wages) and, subsequently, exacerbating income inequality (Arrieta-Ibarra et al., 2018; Posner and Weyl, 2018).

To form a more equitable relationship between the public and technology companies, important stakeholders, i.e. volunteers, the public, and policymakers, need transparent and rigorous evidence about the value of privatized online volunteer work. Without evidence about their work's monetary value, volunteers remain uninformed and disadvantaged when seeking to shape the technologies they co-create with companies. This disadvantage has posed a drag on volunteer productivity and business growth historically, as seen in the class-action lawsuit by AOL moderators in 1999 (Postigo, 2009) and the collective protest by Reddit moderators in 2015 (Matias, 2016). More broadly, opacity in online volunteer work's value hinders the public's ability to address corporate influence on the technological ecosystem that is powered by members of the public (Vincent et al., 2021). Lastly, policymakers, despite making an effort to account for the privatization of online volunteer work in financial regulations (Au-Yeung, 2019; European Commission, 2020, 2017), have not yet had sufficient evidence and knowledge to make pragmatic policy recommendations. In short, assessing the value of online volunteer work is a first step toward supporting volunteers, the public, and policymakers in enabling more equitable power structures in the technology sector.

In this study, we empirically assess the value of a particularly prominent type of online volunteer work—Reddit volunteer moderation. Reddit is one of the most visited websites in the U.S., with fifty-two million daily active users (Patel, 2020), and the website actively plays a role in the public's news consumption (Stoddard, 2015), topical discussions (Gilbert, 2020), and social support for mental health (Chancellor et al., 2019, 2016b; Choudhury and Kiciman, 2017). Reddit is organized into thousands of topical communities, called subreddits. Each subreddit is run by its own volunteer moderators, who make daily decisions about community rules, who may participate, and what content will stay online. Although Reddit moderators have reported the benefits of volunteer moderation models such as independence and tailored community experiences (Chandrasekharan et al., 2018; Gilbert, 2020; Jhaver et al., 2019a; Kiene et al., 2016; Matias, 2019), many also experience frustration about their labor not being recognized and supported by the platform they help to maintain (Gilbert, 2020; Matias, 2019, 2016).

To assess the value of labor contributed by Reddit moderators, we use a novel dataset of private moderator logs ("mod logs") that we collected from 126 communities by working with moderators themselves. This dataset provides more comprehensive coverage of moderation activities than any existing datasets (such as publicly available datasets of removed comments) and allows us to infer the *minimum* amount of time moderators volunteered. Using linear mixed-effect regression, we estimate that the whole volunteer moderator population on Reddit spent *at minimum* 466 hours every day performing moderation actions in 2020. Using the median hourly rate among U.S. commercial content moderators on UpWork ($20/hr), we estimate these labor hours amount to 3.4 million USD a year, equivalent to 3% of Reddit's revenue in 2019.

Our work provides the first empirical estimate of the value of volunteer work that powers Reddit, a highly valued social networking site. In doing so, we contribute a better understanding of online volunteer work's role in technology companies' financial success, and, ultimately, help to inform collective negotiation, public debates, and policy recommendations. Additionally, our ability to draw the estimate is unlocked by our novel method and we discuss how future research can generalize our approach to different types of online volunteer work.

## Related Work

### Online Volunteer Work and Its Impact

Online volunteer work, ranging from open-source projects to peer production to content moderation plays a crucial part in the day-to-day function of prominent computing systems. In recent years, as the output of online volunteer work is repurposed for technological innovation such as language models, online volunteers become a digital labor force that is more important than ever. For example, revolutionary language models such as GPT-3 depend on texts volunteered by Wikipedia editors and Reddit users alike (Brown et al., 2020). Similarly, GitHub's Copilot, a technology that assists programmers in programming is built upon codes published on GitHub. And prominently, many commercial companies and, in particular, cloud computing platforms, benefit from Linux volunteers' work tremendously.

While online volunteer work plays a central role in open-source and commercial computing systems, how to value this work remains unresolved. Researchers have examined how specific outcomes of online volunteer work such as Wikipedia links have benefited technology companies (e.g. (Heald et al., 2015; Piccardi et al., 2021; Vincent et al., 2018)); however, there has not been a way to comprehensively assess the entirety of human labor. Our work takes a first step towards solving this problem by focusing on labor hours, as described in detail below.

### Content Moderation

Content moderation work on Reddit is a prominent case of online volunteer work. Reddit relies on its volunteer moderators to manage thousands of online communities to keep its business viable. These moderators perform a wide range of tasks such as setting up community rules, approving content, removing harmful content, and providing explanations for content removal (e.g. (Gilbert, 2020; Jhaver et al., 2019b)). Moderators also regularly communicate with their peers and community members to discuss and shape community norms (Dosono and Semaan, 2019; Gilbert, 2020). However, much of this work does not leave any publicly visible traces on Reddit (Li et al., 2022). Moderator logs, a type of private data that is only accessible to a subreddit's moderators, provide an opportunity to more comprehensively capture moderation activities than using publicly visible traces such as removed comments. Although mod logs do not capture all moderator activities, they are a step forward in accounting for the invisible part of moderator labor.

The volunteer-driven approach to content moderation is not the only one employed by social platforms; another approach commonly seen in the technology industry is to hire commercial content moderators who moderate content for compensation (Roberts, 2019; Seering, 2020). Compared to the volunteer-driven approach, commercial content moderators have a set of platform guidelines to follow and, therefore, potentially have less independence in managing online communities. However, both groups of moderators perform similar activities; Ruckenstein and Turunen argued that "commercial content moderation has similar aims as community moderation, seeking to support and nurture the online conversation with situated practices" (Ruckenstein and Turunen, 2019).

# Methods

## Data Collection

Estimating the amount of time all Reddit moderators spend on moderation is very difficult because there is no publicly available, comprehensive data about moderator actions and activities. Prior work has inferred moderator activity from the amount of public content removed to approximate overall moderation labor (Chancellor et al., 2016a; Cheng et al., 2015; Lin et al., 2017). However, this removal-based approach underestimates the amount of work volunteer moderators complete because content removal is known to be a fraction of their activities (Gilbert, 2020; Lo, 2018).

To address this major impediment to capturing online volunteer labor, we collected private moderator logs ("mod logs"), which record a wide range of moderator activities in addition to removal actions (see Figure 1 for an example). Mod logs are a list of moderator actions taken on a given subreddit, including information on 1) who took the action, 2) the date and time, 3) the type of action taken, such as approving comments, banning users, and editing the subreddit's Wiki, and 4) to what content or whom the action applies. Mod logs are immutable and updated immediately after moderation actions occur, making them a reliable record of most moderator activities. Despite being the most comprehensive digital record about moderator activities, mod logs do not capture the entirety of the work moderators do and provide an estimate of the *minimum* amount of time spent by moderators. We discuss this limitation below.

Critically, each subreddit's mod logs are only accessible to the subreddit's moderators and are not publicly visible. As such, we gathered mod logs from two sources:

**u/publicmodlogs:** This is a bot account on Reddit that, when added to a subreddit as a moderator, makes all the subreddit's mod logs publicly visible online. The bot was originally developed and used by moderators who wish to make transparent their moderation practices.[1] We included the 84 subreddits that are moderated by u/publicmodlogs and are currently active (i.e. having at least one post and one comment per day).

**Custom Data Collection Bot:** Because the 84 subreddits from u/publicmodlogs are generally dedicated to niche interests, we randomly selected 400 subreddits with the Reddit API's r/random function to sample a diverse set of subreddits. We invited moderators to participate in our research study by sending the mod team a private message (mod mail). During this recruitment, we worked closely with moderators and integrated their feedback on what information in mod logs should be anonymized or omitted during our data collection. Thirty-six subreddits agreed to share their data by adding a data collection bot that we built to their subreddits. Six additional subreddits were recruited through the moderators of the 36 subreddits that also moderated one of those six subreddits. We made our bot's script publicly available online for moderators who are interested in providing feedback on our data collection process.[2] Once added to a subreddit, our bot started collecting its mod logs via the Reddit API. This part of our data collection was reviewed by our Institute Review Board.

Our final dataset consists of mod logs from 126 subreddits (84 subreddits from u/publicmodlogs and 42 subreddits recruited by the research team) for an average of 142 days. These subreddits cover a wide range of topics such as news, politics, humor, and gaming. Table 1 provides information about these subreddits' subscriber count, activity metrics, and data collection span. Although u/publicmodlogs made its affiliated subreddits' mod logs public, it is possible that these subreddits' moderators do not wish to be publicized. As such, to prevent them from being identified, we rounded all subreddits' subscriber counts and activity metrics.

Because we are only interested in human moderators in estimating volunteer hours, we removed automated moderator accounts or bots from the dataset, drawing from methods used in prior work (Jhaver et al., 2019b; Johnson et al., 2016; Warncke-Wang et al., 2013). After removing bot accounts, this dataset contains over 800,000 actions from over 900 human moderators.

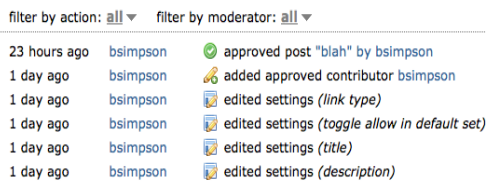

Figure 1: Reddit announced the mod logs feature in 2012 with this screenshot (https://www.reddit.com/r/mod-news/comments/nkj5s/moderators_moderation_log)

|  | Subscriber count in thousands | Daily average post count | Daily average comment count | Data collection span in days |
|---|---|---|---|---|
| *Mean* | 350+ | 70+ | 700+ | 142 |
| *Max* | 15,000+ | 2000+ | 20,000+ | 624 |
| *75%* | 200+ | 40+ | 500+ | 169 |
| *Median* | 50+ | 15+ | 100+ | 167 |
| *25%* | 20+ | 5+ | 20+ | 88 |
| *Min* | 5+ | 1 | 1 | 12 |

Table 1: An overview of our 126 subreddits' subscriber count, activity metrics, and data collection span.

---

[1] https://www.reddit.com/user/publicmodlogs/

[2] https://github.com/hanlinl/modresearch/blob/main/redacted-redditbot.py

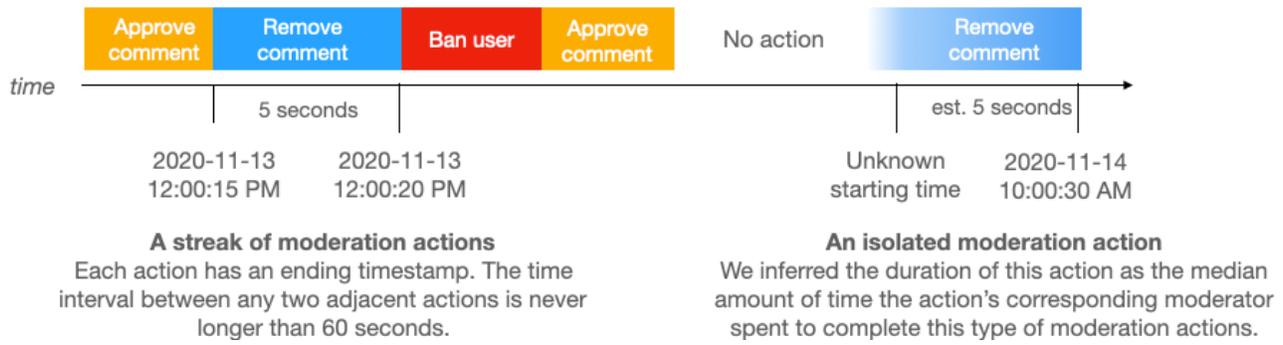

Figure 2: An overview of our approach to assessing how long each moderation action took

### Estimating Moderation Action Duration

To infer how long moderation work took for each moderator based on mod logs, we followed the process from prior work on Wikipedia session analysis of editor activities (Geiger and Halfaker, 2013). Figure 2 provides an overview of our approach. Mod logs only give information on the end timestamp of an action. To estimate how many seconds each action took, we identified "streaks" of actions and calculated how many seconds have passed between each action and its prior action. A "streak" of actions is a series of actions taken sequentially by a subreddit's moderator. Following prior work (Geiger and Halfaker, 2013), we capped the interval between two adjacent actions' end timestamps at 60 seconds[3] or less. Put another way, if 60 seconds have elapsed between two adjacent actions taken by one moderator, these two actions will be classified as belonging to two separate streaks. Because no prior actions exist for isolated actions or the first action in a streak, we assigned each such action the median value in how long the corresponding moderator spent on performing this type of action in general. Finally, we calculated moderation session duration at the moderator level by aggregating our dataset.

### Limitations

Our sample of moderators is not a random sample of the overall moderator population on Reddit. To access one moderator's mod logs from a subreddit, Reddit's API requires our bot to be granted access to mod logs of *all* the moderators on the subreddit. As such, it is not realistic to randomly sample moderators on the site and collect their mod logs, especially given moderators' privacy concerns about sharing access to this type of data. Below, we compared our sample of active moderators with the whole active moderator population. We found that despite statistically significant differences in several activity metrics such as daily distinguished comment count, our sample's means, medians, and standard deviations do not meaningfully deviate from those of the whole active moderator population.

As mentioned earlier, mod logs do not include all moderator activities. For example, replying to moderator mail and deliberation, two types of moderation work reported in prior work (Dosono and Semaan, 2019; Gilbert, 2020) do not appear in mod logs. As a result, our estimate of moderation duration is designed to be *a lower bound estimate*. Although we provide an underestimate, our work serves as a first step towards quantifying this labor and paves the way for future work to comprehensively quantify moderation hours.

### Final Mod Logs Dataset

After collecting mod logs and inferring moderation action duration, we aggregated the dataset to estimate the amount of time moderators spent moderating per day, what we call daily moderation duration. In our sample, daily moderation duration is widely dispersed. This is in part due to a very long tail of inactive moderators – the median in daily moderation duration is 10 seconds (Mean=68 seconds). The number of minutes moderators in our dataset spent every day amounts to 1023 minutes and the top 10% of moderators are responsible for 68% of the daily total, spending between 3 to 40 minutes daily on moderation. The top 20% of moderators spend more than one minute per day and are responsible for 82% of the daily total, which approximately follows the Pareto principle (the 80/20 Rule). For a breakdown of moderation actions, see our in-depth analysis of the makeup of moderator labor using this dataset (Li et al., 2022).

---

[3] We tested multiple values for the threshold we used to separate streaks (30 seconds, 60 seconds, 90 seconds, and 120 seconds). We retained 60 seconds, because the median value in the amount of time elapsed between two adjacent actions began staying stable at this threshold, suggesting that most actions took less than 60 seconds.

|  | Daily moderation duration | Number of subscribers | Daily post count | Daily post count per subscriber | Comment count per post | Number of active moderators |
|---|---|---|---|---|---|---|
| **Number of subscribers** | 0.12*** | | | | | |
| **Daily post count** | 0.26*** | 0.28*** | | | | |
| **Daily post count per subscriber** | 0.06 | -0.16*** | 0.19*** | | | |
| **Comment count per post** | -0.01 | -0.16*** | -0.18*** | -0.24*** | | |
| **Number of active moderators** | 0.21*** | 0.41*** | 0.82*** | -0.08* | -0.13*** | |
| **Daily distinguished comment count moderator** | 0.6*** | 0.27*** | -0.02 | -0.04 | -0.0 | 0.03 |

Table 2: Correlation matrix for main publicly available activity metrics. Note: *p<0.05; ***p<0.001.

|  | Our sample N=378 | | | | | The whole population of active moderators N=21,522 | | | | |
|---|---|---|---|---|---|---|---|---|---|---|
|  | Mean | Median | Std | Max | Min | Mean | Median | Std | Max | Min |
| **Distinguished comment count** | 1.08 | 0.23 | 2.47 | 17.94 | 0.02 | 0.81 | 0.1 | 3.49 | 119.80 | 0.02 |
| **Distinguished post count** | 0.01 | 0 | 0.02 | 0.32 | 0 | 0.01 | 0 | 0.06 | 3.98 | 0 |
| **General comment count** | 6.11 | 2.90 | 9.54 | 87.67 | 0.02 | 6.09 | 2.38 | 11.74 | 250.82 | 0.02 |
| **General post count** | 0.50 | 0.07 | 2.34 | 40.31 | 0 | 0.79 | 0.08 | 4.29 | 229.95 | 0 |
| **Account age in days** | 2248.14 | 2311.63 | 1114.33 | 5403.98 | 180.19 | 2072.25 | 2050.67 | 1208.87 | 5729.53 | 57.86 |
| **Comment karma** | 92454.24 | 38433.5 | 212544.26 | 3080329 | 11 | 61049.88 | 17499.5 | 155825.65 | 4509949 | 1 |
| **Link Karma** | 104273.18 | 12638 | 382020.75 | 5587774 | 1 | 112384.42 | 9681.5 | 714638.92 | 35589509 | 1 |

Table 3: A Comparison of Our Sample and the Whole Active Moderator Population

To provide further insights into daily moderation duration, Table 2 provides the correlation matrix for main publicly available activity metrics. These metrics are from the metadata, posts, and comments collected from Reddit's official API service and the Pushshift Reddit API, a volunteer-led repository of Reddit comments and posts used widely in scientific research (Baumgartner et al., 2020). Most prominently, more active moderators, i.e. moderators who work longer daily, are more likely to leave distinguished comments (comments that are posted by a moderator and publicly marked with a badge icon or a "[M]" label) on their subreddits (spearman's rho=0.60, p<0.001). They are also somewhat likely to be in subreddits with more posts daily (spearman's rho=0.26, p<0.001) and more active moderators (spearman's rho=0.21, p<0 .001).

### Statistics about Hourly Rates for Comparable Commercial Content Moderation Service

To estimate the value of the hours moderators spent on moderation work, we sought to collect statistics about the hourly "wage" or payment rate for comparable paid work. Currently, official wage data sources such as the U.S. Bureau of Labor Statistics do not offer statistics about commercial content moderators, possibly due to this position being a relatively new occupation. To generate an alternative source of comparable wage data, we turned to UpWork, a crowdwork marketplace in which content moderation experts publish their hourly rates to potential clients. We used UpWork because it is a prominent crowdsourcing marketplace and a frequent subject of research, e.g. (Foong et al., 2018; Foong and Gerber, 2021).

We found 160 commercial content moderators on UpWork by searching for keywords related to content moderation and community management services such as "content moderation" and "community manager". We used a new account to minimize personalization in search results. Commercial content moderators' rates varied widely, from $3/hr to $160/hr, and have a long-tailed distribution (Median=$10.0/hr, Mean=$14.6/hr, all in USD). Out of the 160 workers, 31 are located in the United States and their median hourly rate is $20/hr (Mean=$26/hr). The Philippines, where many U.S.-based social media platforms such as YouTube and Facebook employ many commercial content moderators, has the largest number of workers in our dataset, 50, and the median hourly rate is $6/hr (Mean=10/hr) for this population. Reddit moderation work may differ from commercial content moderation services at a granular level; however, they are somewhat comparable given the two roles sharing similar goals and activities (Ruckenstein and Turunen, 2019). As such, the hourly rate for commercial content moderation can serve as a reasonable proxy for volunteer moderation's market value.

### Modeling Moderation Duration

To estimate how many hours of moderation work occurred on Reddit site-wide, we built a linear mixed-effect regression model to extrapolate from our sample of moderators to Reddit's overall moderator population. Specifically, we regressed sampled moderators' daily moderation duration onto their publicly available activity metrics and then applied the model to the overall moderator population.

As mentioned above, our sample is not a truly random sample of moderators. Before proceeding to modeling, we first verified that the moderators in our dataset are at least a somewhat representative sample of the whole active moderator population on Reddit such that findings from our sample can be reasonably generalized. We defined active moderators as moderators who left at least one distinguished comment between November 2020 and January 2021 on the subreddit they moderate. This definition is motivated by the fact that distinguished comment provides high recall for active moderators in our sample: the 378 moderators who have left at least one distinguished comment contributed 97% of the actions in our dataset. We identified 21,522 active moderators on Reddit in 2020 using this technique.

Table 3 shows the comparison between our sample and the population of active moderators using publicly available key activity metrics. The differences in these metrics' means and medians are either insignificant or small—i.e. although the differences are significant, the effect sizes are minimal. However, notably, all two-sample K-S tests reject the null hypothesis that our sample's distributions of these activity metrics follow those of the whole population. Therefore, we expect models derived from our sample would reasonably but imperfectly generalize to the whole population. Put another way, while our sample provides a novel and deep look into moderator activities, this insight requires a moderate sacrifice in sample-population alignment, which is a common limitation in situations in which representative sampling is difficult to implement, e.g. (Killingsworth, 2021).

We regressed daily moderation duration onto publicly available subreddit and moderator activity metrics, using a linear mixed-effect regression model with the 378 active moderators in our sample. We use log-transformed moderation duration as the dependent variable so that the model has normally distributed residuals. The distribution of moderation duration among a subreddit's moderators varies across subreddits. In particular, the Gini index for the 32 subreddits with no fewer than ten human moderators ranges from 0.23 to 0.94 (median = 0.76), suggesting different levels of inequality in workload distribution across subreddits. As such, we chose the best fitting covariance structure to address heterogeneity in residuals (Zuur et al., 2009).

We added a subreddit-specific random intercept to account for the hierarchical structure of our dataset, i.e. moderators being grouped into subreddits. We also added a random slope for distinguished comment count due to the potential variability in its association with daily moderation duration across subreddits. Figure 3 plots the association of the two variables at the dataset level as well as the linear fits between the two metrics for the four subreddits with the largest number of moderators. While distinguished comment count is strongly correlated with daily moderation duration overall (Figure 3 (a)), the relationship between distinguished comment count and moderation duration may be subreddit-specific (Figure 3 (b)). These varied slopes justify a random slope for distinguished comment count per subreddit in our model, which indeed significantly improved our model fit as measured by AIC.

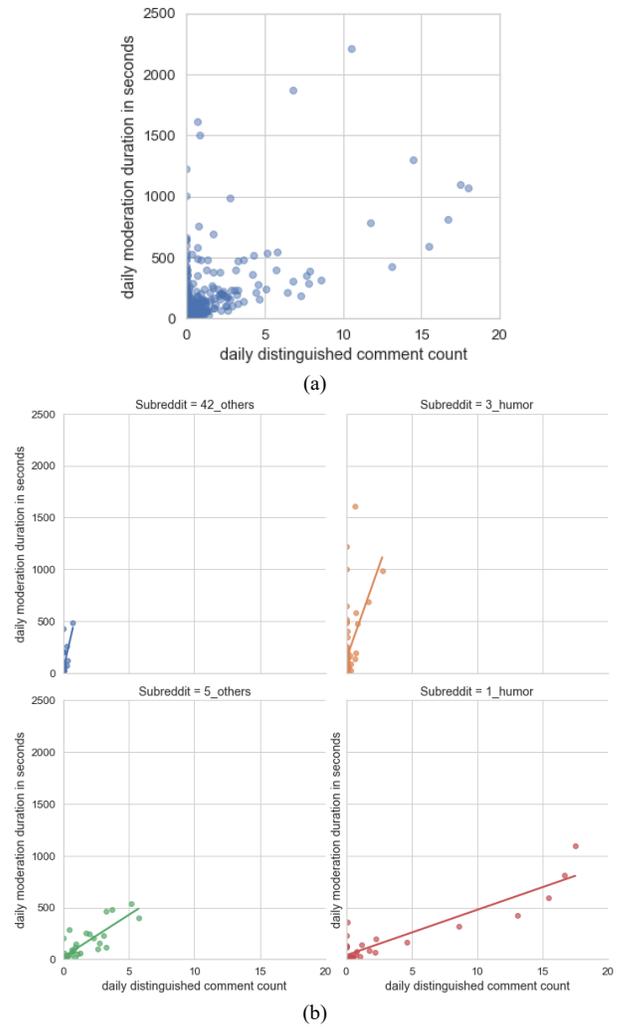

Figure 3: Association between distinguished comment count and moderation duration in the whole dataset (a) and the four subreddits with the largest number of moderators (b)

|  | Estimate |
|---|---|
| **Log(Daily distinguished comment count )** | 0.81 *** |
| **Log(overall comment count / distinguished comment count)** | 0.05 |
| **log(Comment karma )** | 1.24 * |
| **Log(Link karma)** | 0.70 |
| **Account age in days** | -1.28 ** |
| **Log(subreddit daily post count)** | 10.48 *** |
| **Log(Subreddit daily comment count per post)** | 4.45 *** |
| **Subscriber count** | -9.22 *** |
| **NSFW** | 0.24 *** |
| **Log(Number of active mods )** | -0.02 *** |

Table 4: Numeric results of the linear mixed effect regression. *p<0.05, **p<0.01, *** P < 0.001

Table 4 shows the coefficients of the explanatory variables from the model. Controlling for other moderator and subreddit activity metrics and for random variability at the subreddit level, with 1% of increase in distinguished comment count, moderators spend 0.8% more time on their moderation duties. This association points to a promising proxy for future research that seeks to understand and model moderation workload on Reddit. Using a linear regression model with distinguished comment count as the only explanatory variable, we observe that this metric explains 44% of the variance in daily moderation duration across moderators. Future work that seeks to study and identify active moderators may use distinguished comment count as a key indicator.

## Results

Applying our model to all the active moderators on Reddit in 2020, we estimate that moderators spent a total of 466 hours per day performing moderation actions.

As a robustness check for our extrapolation, we calculated the sample's weighted mean in daily moderation duration as a proxy to the population mean. We followed the propensity score weighting approach. We first calculated each moderator's propensity score, i.e. the probability of being included in our sample with a logistic regression model. We then used the inverse of propensity score as each moderator's weight. This weighting process yielded a mean value of 80 seconds per day. This weighted mean corresponds to a sum of 359 to 611 hours by the whole population at the 95% confidence interval, encompassing the point estimate derived from our regression model.

### Applying Hourly Rates

The value of the 466 labor hours is contingent on the market rate for content moderation. We take two approaches for our estimation. In our first approach, we assume that companies in the market for content moderation services always hire workers who charge the least. In our case, the 59 such workers (466/8, assuming that each worker can work eight hours a day) charge $3/hr to $12/hr with a mean of $8/hr. As a result, this approach leads to an estimate of 1.4 million USD for the 466×365 estimated moderation hours, equivalent to 1% of Reddit's revenue in 2019 (120 million USD).

In our second approach, we assume that companies looking to hire content moderators offer a fixed rate. We consider several rates from the UpWork dataset in addition to the $15 hourly wage strongly advocated by scholars and crowdworers (Rolf, 2015; Whiting et al., 2019) and the U.S. federal minimum wage (given the U.S.'s status as the primary market for Reddit (Statista, 2021)). We report all the estimates in Table 5. The yearly value of Reddit volunteer moderators' labor is 3.4 million USD (3% of Reddit's revenue in 2019) if calculated with the median rate of U.S.-based UpWork workers and 4.4 million USD (4% of Reddit's revenue in 2019) if calculated with the mean. Additionally, we consider the Philippines, the country with a prominent labor force of content moderators (Roberts, 2019); the amount of work volunteer moderators completed on Reddit is worth 1 million USD using the median rate of Philippines-based workers and 1.9 million USD using the mean.

|  | Hourly rate | Annual worth (percentage of revenue) |
|---|---|---|
| **UpWork Global median** | $10 | 1.7 million USD (1.4%) |
| **UpWork Global mean** | $14.6 | 2.5 million USD (2.1%) |
| **UpWork U.S. median** | $20 | 3.4 million USD (2.8%) |
| **UpWork U.S. mean** | $26 | 4.4 million USD (3.7%) |
| **UpWork Philippines median** | $6 | 1.0 million USD (0.8%) |
| **UpWork Philippines mean** | $11 | 1.9 million USD (1.6%) |
| **$15 Minimum Wage** | $15 | 2.6 million USD (2.1%) |
| **U.S. Federal minimum wage** | $7.25 | 1.2 million USD (1.0%) |

Table 5: the worth of volunteer moderators' labor calculated in various rates

## Discussion

Our work quantifies the value of labor subsidy volunteer moderators contribute to Reddit, a highly valued technology company. There exist many other prominent, highly valued technology companies and products that rely on volunteer labor, such as Google Maps, Yelp, and Facebook Groups. In this section, we first explore how our results can assist important stakeholders—online volunteers, the public, and policymakers—in ensuring online volunteers are adequately supported and recognized by these companies. We then discuss how future research may further improve this estimate.

### Implications for Online Volunteers

Moderators have already voiced their displeasure over poor support for their labor. In 2015, Reddit moderators made thousands of subreddits inaccessible to the public in a protest against inadequate tooling and administrative support - this protest blocked much public web traffic to the site (Matias, 2016). Historically, tensions between volunteers and companies have spilled into legal disputes. In the late 1990s and early 2000s, America Online (AOL) moderators filed a class-action lawsuit to dispute the company's management of moderators (Postigo, 2009).

Our estimates highlight potential opportunities for online volunteers and companies to form a better, more sustainable relationship that can maintain the health and vibrancy of the technologies they co-created like Reddit. Our study quantified the important value Reddit moderators bring to the company, and volunteer moderators could highlight this value in their conversation with Reddit to advocate for resources needed to successfully manage online communities. For example, as research and prior historical examples have shown that Reddit's existing moderation tools fail to support volunteer moderators' work (Matias, 2019; Postigo, 2009; Seering et al., 2019), moderators could use the value we describe here as a talking point to demand software engineering efforts that is equivalent to their collective volunteer hours to improve these tools.

More broadly, knowing the amount of this labor subsidy they supply to Reddit can help volunteer moderators to advocate more strongly for decision-making power in the platform's day-to-day operation such as updating site-wide content policies and division of responsibilities between themselves and the site (Matias, 2019). Our finding on the long-tailed distribution of moderation work suggests that a small share of moderators might have particularly strong negotiation power. This raises the question for future work about how (or whether) to ensure that any collective negotiation with Reddit is representative of the diversity of Reddit moderators rather than being driven by a few active moderators.

Our work also raises interesting questions about how Reddit may react to moderators' protests, such as making their subreddits private or quitting. What if Reddit decides to hire commercial content moderators rather than spending time and resources addressing volunteer moderators' concerns? Given volunteer content moderators' close connection with communities and in-depth knowledge about community dynamics, it is unlikely for Reddit to replace volunteer content moderators altogether. However, some subreddits' moderators have expressed interest in asking Reddit to hire commercial content moderators to supplement their labor. A fruitful area of research would be exploring how to bring volunteer and commercial content moderators together to manage online communities.

**Implications for the Public**

By measuring online volunteer work's role in supporting businesses like Reddit, our work can better inform the public of ways to meaningfully shape the technology landscape. Currently, as technology companies that rely on volunteer labor gain more and more power in the technology realm, there exists evidence of their business development's negative impact on the public, e.g. monopolistic practices (Kim and Luca, 2019; Li and Hecht, 2020) and lack of transparency in data use (Sadowski et al., 2021). By making explicit technology companies' dependence on volunteer labor, our work points to an opportunity for the public to collaborate with online volunteers to mitigate companies' influence over the technological ecosystem. Currently, online volunteers and, more broadly, members of the public supply valuable time, data, and knowledge to for-profit technologies such as social media platforms and rating systems but have little say over how these technologies are designed and developed. Our study highlights a potential direction that the public may take to mitigate this power imbalance. For example, the public may join online volunteers' in stopping their use of a technology or migrating to competing technologies—what Vincent et al. coined as "data leverage" (Vincent et al., 2021), to directly divert the crucial data and labor away from certain companies.

**Implications for Policymakers**

Finally, our work can assist policymakers in drafting regulations that can account for unpaid labor subsidies for for-profit technology companies. Economists have noted the sector's declining labor share and exacerbating income inequality (Arrieta-Ibarra et al., 2018; Brynjolfsson and McAfee, 2014; Posner and Weyl, 2018). Our method that estimates online volunteer work's value based on the work's comparable market rate could be adapted to different privatized online volunteer work, ranging from user-generated ratings to image labels. Previously, the monetary value of online volunteer work was difficult to estimate in part due to the complexity and opacity of the large, for-profit technologies it supports. For example, how much ad revenue volunteer moderators bring to Reddit when they remove a harmful post remains nebulous and may require sophisticated experiment design if at all possible. Our method is generalizable to other types of online volunteer work (see more details in Future Work) and could help policymakers understand how much value technology companies benefit from online volunteer work (Au-Yeung, 2019; European Commission, 2017). To more equitably distribute the profits of volunteer-powered technologies, just as tax assessors evaluate a property's market value to determine its owner's tax bill, policymakers could start assessing technology companies' "volunteer-dependence" tax based on the amount of online volunteer work calculated in a similar fashion to this paper.

Policymakers may also fund third-party "volunteer labor auditors" that conduct independent time studies to advocate for union members. Such entities could collect and analyze time logs from volunteers while preserving their privacy. Their findings would then assist volunteers in collective negotiation with companies that benefit from this labor.

## Reflection on Data Collection

Our work is made possible due to the generosity of volunteer moderators who shared their mod logs and the openness of subreddits whose mod logs are made public. Given the sensitive and private nature of our data, we took special caution in our data collection, storage, and analysis. For those subreddits we recruited ourselves, we made our data collection script accessible to moderators so they could directly see what information we would anonymize and collect.

In accordance with our IRB protocol, we cannot make mod logs we collected through our own recruitment public because mod logs contain extremely sensitive information about moderators' behaviors such as who removed what content. During recruitment, several moderators contacted us to confirm that only our research team would have access to their mod logs. Even if we anonymized all target links and account names, other information such as timestamps and action type may still risk moderators being identified.

For the mod logs we collected from u/publicmodlogs, we cannot publish them because the mod logs are no longer publicly available. u/publicmodlogs only publishes the past three months' mod logs from its affiliated subreddits. By the time of this writing, all the mod logs we collected from u/publicmodlogs are no longer accessible online. We respect this setting to preserve the integrity of what moderators may have agreed to in adding u/publicmodlogs to their subreddits.

## Future Work

Future work could extend our study in at least three directions. First, although our dataset is the most comprehensive dataset about moderation work on Reddit, our estimate is conservative. This estimate could be improved with more tracking of moderator behaviors. Moderator logs do not include time spent on untraced activities like responding to moderator mail, debating about moderation decisions on other platforms like Discord or Slack, and developing moderation bots. Additionally, our estimate only considers moderators of public Reddit communities because data about private communities' moderators is not accessible. Therefore, our estimate of moderator labor is a floor of the true amount of time Reddit's moderator population spends moderating on the site. Future work may enhance our estimate by analyzing moderator activities across tools and platforms from both public and private communities.

Second, the hourly rate of commercial content moderation service from UpWork may underestimate moderation work's worth. Prior research on Amazon Mechanical Turk shows that monopsony power drives down crowd workers' wages (Dube et al., 2020), which may also occur on UpWork to a lesser degree. Future work may collect more wage data to improve our estimate. Nonetheless, this approach can provide an important starting point about the value of online volunteer work.

Third, our analytical method may be generalized to other types of online volunteer work to understand the amount of "labor subsidy" online volunteers supply to additional for-profit technology companies. Future work could explore the value of the labor hours online volunteers spend on writing reviews for products and services, providing implicit and explicit feedback for intelligent models' output, and producing answers on Q&A websites. For example, to assess the monetary value of the ratings volunteers provided on platforms such as Google Maps, one could first construct a reasonably representative sample of online volunteers and conduct a large-scale data collection to infer the hours they spent. Then researchers could model these volunteers' hours with publicly available user metrics such as years active and number of ratings written and use the model to extrapolate to the whole population for an estimate of the total volunteer hours. Finally, since providing ratings is a type of crowdsourcing task on Amazon Mechanical Turk, researchers could use the corresponding wage rate to estimate the monetary value of the estimated volunteer hours.

## Conclusion

Using Reddit moderation as a case study, we estimate the monetary value of the online volunteer work completed by all Reddit volunteer moderators. This estimate may assist companies, volunteers, and policymakers in proactively upholding the volunteer-driven business model, the foundation of many successful technology companies and technological advancements. Our estimate is enabled by a novel method that projects online volunteer work's value based on equivalent, commercial services, and has the potential to be adapted for volunteer work beyond content moderation.

## Acknowledgments

The authors would like to thank Sanhita Sengupta from the Statistical Consulting Center at the University of Minnesota Twin Cities, Darren Gergle and Nicholas Diakopoulos from Northwestern University, and Charles Kiene from the Community Data Science Collective for providing feedback on this work. The authors would also like to thank Nicholas Vincent from Northwestern University for offering suggestions for Reddit data collection. This study is made possible by the mod logs contributed by Reddit moderators.

## References

Arrieta-Ibarra, I.; Goff, L.; Jiménez-Hernández, D.; Lanier, J.; Weyl, E.G., 2018. Should We Treat Data as Labor? Moving Beyond "Free." AEA Papers and Proceedings 108, 38–42. https://doi.org/10.1257/pandp.20181003

Au-Yeung, A., 2019. California Wants To Copy Alaska And Pay People A 'Data Dividend.' Is It Realistic? [WWW

Document]. Forbes. URL https://www.forbes.com/sites/angelauyeung/2019/02/14/california-wants-to-copy-alaska-and-pay-people-a-data-dividend--is-it-realistic/ (accessed 4.20.21).

Baumgartner, J.; Zannettou, S.; Keegan, B.; Squire, M.; Blackburn, J., 2020. The Pushshift Reddit Dataset. ICWSM 14, 830–839.

Brown, T.B.; Mann, B.; Ryder, N.; Subbiah, M.; Kaplan, J.; Dhariwal, P.; Neelakantan, A.; Shyam, P.; Sastry, G.; Askell, A.; Agarwal, S.; Herbert-Voss, A.; Krueger, G.; Henighan, T.; Child, R.; Ramesh, A.; Ziegler, D.M.; Wu, J.; Winter, C.; Hesse, C.; Chen, M.; Sigler, E.; Litwin, M.; Gray, S.; Chess, B.; Clark, J.; Berner, C.; McCandlish, S.; Radford, A.; Sutskever, I.; Amodei, D., 2020. Language Models are Few-Shot Learners. arXiv:2005.14165 [cs].

Brynjolfsson, E.; McAfee, A., 2014. The Second Machine Age: Work, Progress, and Prosperity in a Time of Brilliant Technologies. W. W. Norton & Company.

Chancellor, S.; Lin, Z.; Goodman, E.L.; Zerwas, S.; De Choudhury, M., 2016a. Quantifying and Predicting Mental Illness Severity in Online Pro-Eating Disorder Communities, in: Proceedings of the 19th ACM Conference on Computer-Supported Cooperative Work & Social Computing, CSCW '16. Association for Computing Machinery, New York, NY, USA, pp. 1171–1184. https://doi.org/10.1145/2818048.2819973

Chancellor, S.; Mitra, T.; De Choudhury, M., 2016b. Recovery Amid Pro-Anorexia: Analysis of Recovery in Social Media. ACM Press, pp. 2111–2123. https://doi.org/10.1145/2858036.2858246

Chancellor, S.; Nitzburg, G.; Hu, A.; Zampieri, F.; De Choudhury, M., 2019. Discovering Alternative Treatments for Opioid Use Recovery Using Social Media, in: Proceedings of the 2019 CHI Conference on Human Factors in Computing Systems. Presented at the CHI '19: CHI Conference on Human Factors in Computing Systems, ACM, Glasgow Scotland Uk, pp. 1–15. https://doi.org/10.1145/3290605.3300354

Chandrasekharan, E.; Samory, M.; Jhaver, S.; Charvat, H.; Bruckman, A.; Lampe, C.; Eisenstein, J.; Gilbert, E., 2018. The Internet's Hidden Rules: An Empirical Study of Reddit Norm Violations at Micro, Meso, and Macro Scales. Proceedings of the ACM on Human-Computer Interaction 2, 1–25. https://doi.org/10.1145/3274301

Cheng, J.; Danescu-Niculescu-Mizil, C.; Leskovec, J., 2015. Antisocial Behavior in Online Discussion Communities 10.

Choudhury, M.D.; Kiciman, E., 2017. The Language of Social Support in Social Media and Its Effect on Suicidal Ideation Risk. Proceedings of the International AAAI Conference on Web and Social Media.

Dosono, B.; Semaan, B., 2019. Moderation Practices as Emotional Labor in Sustaining Online Communities: The Case of AAPI Identity Work on Reddit, in: Proceedings of the 2019 CHI Conference on Human Factors in Computing Systems, CHI '19. Association for Computing Machinery, Glasgow, Scotland Uk, pp. 1–13. https://doi.org/10.1145/3290605.3300372

Dube, A.; Jacobs, J.; Naidu, S.; Suri, S., 2020. Monopsony in Online Labor Markets. American Economic Review: Insights 2, 33–46. https://doi.org/10.1257/aeri.20180150

European Commission, 2020. Proposal for a Regulation on European data governance (Data Governance Act) | Shaping Europe's digital future [WWW Document]. URL https://digital-strategy.ec.europa.eu/en/library/proposal-regulation-european-data-governance-data-governance-act (accessed 4.20.21).

European Commission, 2017. Fair Taxation of the Digital Economy [WWW Document]. Taxation and Customs Union - European Commission. URL https://ec.europa.eu/taxation_customs/business/company-tax/fair-taxation-digital-economy_en (accessed 3.16.21).

Foong, E.; Gerber, E., 2021. Understanding Gender Differences in Pricing Strategies in Online Labor Marketplaces, in: Proceedings of the 2021 CHI Conference on Human Factors in Computing Systems. Association for Computing Machinery, New York, NY, USA.

Foong, E.; Vincent, N.; Hecht, B.; Gerber, E.M., 2018. Women (Still) Ask For Less: Gender Differences in Hourly Rate in an Online Labor Marketplace. Proc. ACM Hum.-Comput. Interact. 2. https://doi.org/10.1145/3274322

Geiger, R.S.; Halfaker, A., 2013. Using edit sessions to measure participation in wikipedia, in: Proceedings of the 2013 Conference on Computer Supported Cooperative Work, CSCW '13. Association for Computing Machinery, San Antonio, Texas, USA, pp. 861–870. https://doi.org/10.1145/2441776.2441873

Gilbert, S.A., 2020. "I run the world's largest historical outreach project and it's on a cesspool of a website." Moderating a Public Scholarship Site on Reddit: A Case Study of r/AskHistorians. Proc. ACM Hum.-Comput. Interact. 4, 019:1-019:27. https://doi.org/10.1145/3392822

Heald, P.; Erickson, K.; Kretschmer, M., 2015. The Valuation of Unprotected Works: A Case Study of Public Domain Images on Wikipedia. Harv. J. L. & Tech. 29, 1.

Jhaver, S.; Birman, I.; Gilbert, E.; Bruckman, A., 2019a. Human-Machine Collaboration for Content Regulation: The Case of Reddit Automoderator. ACM Transactions on Computer-Human Interaction 26, 1–35. https://doi.org/10.1145/3338243

Jhaver, S.; Bruckman, A.; Gilbert, E., 2019b. Does Transparency in Moderation Really Matter? User Behavior After Content Removal Explanations on Reddit. Proc. ACM Hum.-Comput. Interact. 3, 150:1-150:27. https://doi.org/10.1145/3359252

Johnson, I.L.; Lin, Y.; Li, T.J.-J.; Hall, A.; Halfaker, A.; Schöning, J.; Hecht, B., 2016. Not at Home on the Range: Peer Production and the Urban/Rural Divide, in: Proceedings of the 2016 CHI Conference on Human Factors in Computing Systems, CHI '16. ACM, New York, NY, USA, pp. 13–25. https://doi.org/10.1145/2858036.2858123

Kiene, C.; Monroy-Hernández, A.; Hill, B.M., 2016. Surviving an "Eternal September": How an Online Community Managed a Surge of Newcomers, in: Proceedings of the 2016 CHI Conference on Human Factors in Computing Systems. Presented at the CHI'16: CHI Conference on Human Factors in Computing Systems, ACM, San Jose California USA, pp. 1152–1156. https://doi.org/10.1145/2858036.2858356

Killingsworth, M.A., 2021. Experienced well-being rises with income, even above $75,000 per year. PNAS 118. https://doi.org/10.1073/pnas.2016976118

Kim, H.; Luca, M., 2019. Product Quality and Entering Through Tying: Experimental Evidence. Management Science 65, 596–603. https://doi.org/10.1287/mnsc.2018.3246

Li, H.; Hecht, B.; Chancellor, S., 2022. All That's Happening behind the Scenes: Putting the Spotlight on Volunteer Moderator Labor in Reddit. ICWSM.

Li, H.; Hecht, B., 2020. 3 Stars on Yelp, 4 Stars on Google Maps: A Cross-Platform Examination of Restauration Ratings. Proceedings of the ACM on Human-Computer Interaction 2.

Li, H.; Vincent, N.; Tsai, J.; Kaye, J.; Hecht, B., 2019. How do people change their technology use in protest?: Understanding "protest users." Proceedings of the ACM on Human-Computer Interaction 3, 87.

Lin, Z.; Salehi, N.; Yao, B.; Chen, Y.; Bernstein, M.S., 2017. Better When It Was Smaller? Community Content and Behavior After Massive Growth 10.

Lo, C., 2018. When All You Have is a Banhammer: The Social and Communicative Work of Volunteer Moderators (Thesis).

Matias, J.N., 2019. The Civic Labor of Volunteer Moderators Online. Social Media + Society 5, 2056305119836778. https://doi.org/10.1177/2056305119836778

Matias, J.N., 2016. Going Dark: Social Factors in Collective Action Against Platform Operators in the Reddit Blackout, in: Proceedings of the 2016 CHI Conference on Human Factors in Computing Systems, CHI '16. ACM, New York, NY, USA, pp. 1138–1151. https://doi.org/10.1145/2858036.2858391

Patel, S., 2020. Reddit Claims 52 Million Daily Users, Revealing a Key Figure for Social-Media Platforms. Wall Street Journal.

Piccardi, T.; Redi, M.; Colavizza, G.; West, R., 2021. On the Value of Wikipedia as a Gateway to the Web, in: Proceedings of the Web Conference 2021. Association for Computing Machinery, New York, NY, USA, pp. 249–260.

Posner, E.A.; Weyl, E.G., 2018. Radical markets: Uprooting capitalism and democracy for a just society. Princeton University Press.

Postigo, H., 2009. America Online volunteers: Lessons from an early co-production community. International Journal of Cultural Studies 12, 451–469. https://doi.org/10.1177/1367877909337858

Roberts, S.T., 2019. Behind the Screen, Behind the Screen. Yale University Press.

Rolf, D., 2015. The Fight for $15: The Right Wage for a Working America. New Press, The.

Ruckenstein, M.; Turunen, L.L.M., 2019. Re-humanizing the platform: Content moderators and the logic of care: New Media & Society. https://doi.org/10.1177/1461444819875990

Sadowski, J.; Viljoen, S.; Whittaker, M., 2021. Everyone should decide how their digital data are used — not just tech companies. Nature 595, 169–171. https://doi.org/10.1038/d41586-021-01812-3

Seering, J., 2020. Reconsidering Self-Moderation: the Role of Research in Supporting Community-Based Models for Online Content Moderation. Proc. ACM Hum.-Comput. Interact. 4, 107:1-107:28. https://doi.org/10.1145/3415178

Seering, J.; Wang, T.; Yoon, J.; Kaufman, G., 2019. Moderator engagement and community development in the age of algorithms. New Media & Society 21, 1417–1443. https://doi.org/10.1177/1461444818821316

Statista, 2021. Reddit: traffic by country [WWW Document]. Statista. URL https://www.statista.com/statistics/325144/reddit-global-active-user-distribution/ (accessed 1.11.22).

Stoddard, G., 2015. Popularity Dynamics and Intrinsic Quality in Reddit and Hacker News 10.

Terranova, T., 2000. Free Labor: Producing Culture for the Digital Economy. Social Text 18. https://doi.org/10.1215/01642472-18-2_63-33

Vincent, N.; Johnson, I.; Hecht, B., 2018. Examining Wikipedia with a broader lens: Quantifying the value of Wikipedia's relationships with other large-scale online communities, in: Proceedings of the 2018 CHI Conference on Human Factors in Computing Systems, CHI '18. ACM, New York, NY, p. 566:1-566:13. https://doi.org/10.1145/3173574.3174140

Vincent, N.; Li, H.; Tilly, N.; Chancellor, S.; Hecht, B., 2021. Data Leverage: A Framework for Empowering the Public in its Relationship with Technology Companies, in: Proceedings of the 2021 ACM Conference on Fairness, Accountability, and Transparency. Association for Computing Machinery, New York, NY, USA, pp. 215–227.

Warncke-Wang, M.; Cosley, D.; Riedl, J., 2013. Tell me more: an actionable quality model for Wikipedia, in: Proceedings of the 9th International Symposium on Open Collaboration, WikiSym '13. Association for Computing Machinery, New York, NY, USA, pp. 1–10. https://doi.org/10.1145/2491055.2491063

Whiting, M.E.; Hugh, G.; Bernstein, M.S., 2019. Fair Work: Crowd Work Minimum Wage with One Line of Code. Proceedings of the AAAI Conference on Human Computation and Crowdsourcing 7, 197–206.

Zuur, A.F.; Ieno, E.N.; Walker, N.J.; Saveliev, A.A.; Smith, G.M., 2009. Dealing with Heterogeneity, in: Zuur, A.F., Ieno, E.N., Walker, N., Saveliev, A.A., Smith, G.M. (Eds.), Mixed Effects Models and Extensions in Ecology with R, Statistics for Biology and Health. Springer, New York, NY, pp. 71–100. https://doi.org/10.1007/978-0-387-87458-6_4